\documentclass[12pt]{iopart}
\usepackage{amssymb}
\usepackage{graphicx}
\usepackage{epsfig}
\usepackage{color}

\begin{document}

\title{Statistical mixing and aggregation in Feller diffusion}

\author{C Anteneodo$^1$ and S M Duarte Queir\'{o}s$^2$}

\address{$^1$ Department of Physics, PUC-Rio and \\
National Institute of Science and Technology for Complex Systems \\
Rua Marqu\^es de S\~ao Vicente 225, G\'avea, CEP 22453-900 RJ, Rio de
Janeiro, Brazil}
\address{$^2$ Unilever R\&D Port Sunlight \\
Quarry Road East, Wirral, CH63 3JW UK}

\ead{celia@fis.puc-rio.br}

\begin{abstract}
We consider Feller mean-reverting square-root diffusion, which has been
applied to model a wide variety of processes with linearly state-dependent
diffusion, such as stochastic volatility and interest rates in finance, and
neuronal and populations dynamics in natural sciences.
We focus on the statistical mixing (or superstatistical)
process in which the parameter related to the mean value can fluctuate --- a plausible mechanism for the
emergence of heavy-tailed distributions. We obtain analytical results for the
associated probability density function (both stationary and time
dependent), its correlation structure and aggregation properties. Our
results are applied to explain the statistics of stock traded volume at
different aggregation scales.

\end{abstract}

\vspace{2pc}
\maketitle

\section{Introduction}

Complex non-equilibrium phenomena have lured the attention of a large part
of the physical community in recent years. Despite the knotty character
of this type of systems, by applying particular techniques, it has been
possible to make a physical characterization of their leading properties.
Among these techniques, let us mention the ``superstatistical'' approach. It
originated from the observation, in the context of nuclear collisions, that
deviations from the standard (exponential) probability density function
(PDF) could be explained through the fluctuations of an inner parameter, in
that case, the cross section~\cite{wilk}. Because a statistics is made up of
another statistics, this approach became known as superstatistics, or
statistics of statistics. Therafter, it was further generalised, endowed
with a statistical mechanics interpretation~\cite{beck-cohen} and widely
applied since then. In fact, for many systems, it is realistic to consider
that some of the characteristic parameters may be not strictly constant, but
instead fluctuant, either in time or in space, according to a specific PDF,
in a scale much larger than the primary stochastic process.
The superstatistical approach has been quite successful in accounting for
observations in fluid turbulence~\cite{bodenschatz}, physiology~\cite{beck},
human activities~\cite{soccer}, ecology~\cite{ecology}, and also in finance~%
\cite{bouchaud,ca-vol,smdq-vol}, amid many others. Let us also point out
that, in economics and social sciences, albeit \textit{ad-hoc}, statistical
mixtures have been taken into account for some decades~\cite{mixtures}.

In this work, we consider as primary process the one given by the stochastic
differential equation (SDE)
\begin{equation}
dx=-\gamma \left[ x-\theta \right] \,dt+\delta \,\sqrt{x}\,dW_{t},\quad
\left( x\geq 0\right) ,  \label{sde}
\end{equation}
where $W_{t}$ represents a standard Wiener process, with unitary variance,
and $\gamma $, $\theta $ and $\delta $ are positive real parameters. This
SDE, first studied by Feller~\cite{feller}, is well-known in mathematical
finance. It was employed by Cox, Ingersoll and Ross to model short-term
interest rates~\cite{CIR} and later became popular in mimicking stochastic
volatility, like in the Heston model for price dynamics~\cite{heston}.
Mean-reverting square-root diffusion has also been considered in other
contexts, such as in modelling neural spiking~\cite{spiking} or in problems
of biological diffusion~\cite{bio}.

Still in the context of finance, in the first approximation, Eq.~(\ref{sde}) describes the dynamics
of share trading volumes, although the tails of
the empirical distributions deviate from the steady solution associated with
the SDE (\ref{sde}),
\begin{equation}
P_s(x)=\mathcal{N} \,x^{2\gamma\theta/\delta^2-1 }\exp(-2\gamma x/\delta^2)\,,
\end{equation}
which is the Gamma PDF~\cite{feller}, with $\mathcal{N}$ a normalization
constant. Fluctuations that can explain the observed deviations
within the superstatistical mixing framework~\cite{smdq-vol}
have been detected in the parameter directly related to the mean value~\cite{ca-vol}. The resulting
PDF, known as $q$-Gamma, is a generalization of the Gamma distribution that
can be cast into the form of the $F$-distribution and which basically turns the
exponential tail into a power-law one. It has been shown to be in excellent
agreement with empirical observations at different granularity timescales
(from 1 min to days)~\cite{ca-vol,smdq-vol,obt,volumes}.

The study of complex systems often encompasses the analysis of the
probability function of the addition of stochastic observables, mostly to
appraise the hypothesis of scale invariance. Precisely, in respect of this, the
fact that the description of empirical volume PDFs in terms of $q$-Gamma distributions
applies at different aggregation scales is particularly interesting,
especially because, unlike  the Gamma distribution, the $q$-Gamma is
neither closed under convolution nor correlations can be fully neglected.

To understand these observations motivates the present work.
Although our initial motivation comes from an econophysical problem,
the present results may be of interest for a wider scenario where linear
diffusion applies, as soon as parameter fluctuations are ubiquitous.

The manuscript is organized as follows. We first summarize the pertinent
results related to the Fokker-Planck equation (FPE) associated to the
stochastic process (\ref{sde}). We apply these results to obtain the PDFs
resulting from an accumulation process. Thereafter, we apply the statistical
composition procedure, in which the reverting mean is the fluctuating
parameter. We obtain joint distributions that us allow to characterize the
correlation structure as well as the aggregation properties. Finally, we
apply the analytical results to interpret the granular features of
real time-series of trading volumes.

\section{Primary process}

The forward Fokker-Planck equation (FPE) associated to Eq.~(\ref{sde}), for
the conditional probability $P\equiv P(x,t|x^{\prime },t^{\prime })$, reads
\begin{equation}
\frac{\partial P}{\partial t}=\frac{\partial }{\partial x}\left( \gamma [
x-\theta ] P\right) +\frac{1}{2}\frac{\partial^{2}}{\partial x^{2}} ( \delta
^{2}\,x\,P ) \,.  \label{fpe}
\end{equation}
The propagator of this FPE has been obtained by Feller~\cite{feller}. Since
then, the time-dependent solution has been systematically overlooked. For
instance, in stochastic volatility models like the Heston model and its
variants, the price is the quantity of interest~\cite{heston,yakovenko},
while the volatility, modeled by Eq.~(\ref{sde}), is only an auxiliary
quantity. Wherefore, it is usually integrated out by considering a
distribution of the (initial) volatility equal to the stationary solution.
Thence, if we want to go further, we must look for the time-dependent
solutions. With that goal in mind, we summarize the procedure for the
obtention of the propagator, first presented by Feller~\cite{feller}, that
embodies the definitions and the partial results which are going to be
useful in following sections.

Laplace transforming Eq.~(\ref{fpe}), one gets
\begin{equation}
\frac{\partial \tilde{P}}{\partial t} =-\left(\gamma +\frac{\delta ^{2}}{2}w
\right)\frac{\partial \tilde{P}}{\partial \,w} -\gamma \,\theta \,w\,\tilde{P
}\,,  \label{fp-lagrange}
\end{equation}
where $\tilde{P}\equiv \tilde{P}(w,t|x^{\prime },t^{\prime })$. With the
initial condition $P\left( x,t^{\prime }|x^{\prime },t^{\prime }\right)
=\delta \left( x-x^{\prime }\right)$, whose Laplace transform is $\tilde{P}
\left( w,t^{\prime }|x^{\prime },t^{\prime }\right) =\exp(-w\,x^\prime)$, the solution
of Eq. (\ref{fp-lagrange}), that can be obtained by the method of
characteristics, is
\begin{equation}
\tilde{P}\left( w,t|x^{\prime },t^{\prime }\right) =\frac{\exp(-\frac{Aw}{
1+Bw})}{(1+Bw)^{\beta }} =\frac{\exp(-\frac{A}{B}[1-\frac{1}{1+Bw}] )}{
\left( 1+Bw\right)^{\beta }}\,,  \label{Ltsol}
\end{equation}
where we have defined,
\begin{eqnarray*}
\Theta& \equiv \exp(-\gamma [t-t^{\prime }]), \\
A& \equiv x^{\prime }\Theta, \\
B& \equiv B_{0}[1-\Theta]\equiv \frac{\delta^{2}}{2\gamma }[1-\Theta], \\
\beta & \equiv \frac{2\gamma \theta }{\delta^{2}}=\frac{\theta }{B_{0}}\,.
\end{eqnarray*}
In the long-time limit $\Delta t\equiv (t-t)^{\prime }>>1/\gamma $
(hence, $\Theta\rightarrow 0$), Eq.~(\ref{Ltsol}) becomes
\begin{equation}
\tilde{P}(w,t|x^{\prime },t^{\prime })=\frac{1}{\left( 1+B_{0}w\right)
^{\beta }}\,,  \label{limitLt}
\end{equation}
whose inverse Laplace transform gives us the steady solution in $x$ space,
that reads
\begin{equation}
P_{s}(x)=\mathcal{L}^{-1} \Bigl(\frac{1}{\left( 1+B_{0}w\right) ^{\beta }}
\Bigr) =\frac{ x^{\beta -1} \exp(- \frac{x}{B_0} )}{B_{0}^{\beta }\Gamma
(\beta) }\,,  \label{limitP}
\end{equation}
for nonnegative $x$, and zero otherwise. This is the Gamma (or Erlang)
distribution, $\Gamma _{\beta ,B_{0}}$~\cite{cooper}.

For any $\Delta t$, the conditional PDF $P(x,t|x^{\prime},t^{\prime})$ can
be obtained by first expanding the exponential in Eq.~(\ref{Ltsol}) and then
performing the mappings $\beta\to n+\beta$ and $B_0 \to B$ in
Eq.~(\ref{limitP}). That is,
\begin{eqnarray}
&&P(x,t|x^{\prime},t^{\prime}) =\mathcal{L}^{-1} \bigl( \tilde{P}
(w,t|x^{\prime},t^{\prime}) \bigr)   \\
&=&\sum_{n\geq0}\frac{\exp(-\frac{A}{B})\bigl( \frac{A}{B}\bigr)^{n}}{n!}
\mathcal{L}^{-1} \Bigl( \frac{1}{\left( 1+Bw \right)^{n+\beta}} \Bigr)
 \\
& =& \sum_{n\geq0}\frac{\exp(-\frac{A}{B})\bigl( \frac{A}{B}\bigr)^{n}} {n!}
\frac{x^{n+\beta-1}\exp(-\frac{x}{B}) }{B^{n+\beta}\Gamma( n+\beta) }
\\
& =& \frac{x^\frac{\beta-1}{2} \exp(-\frac{A+x}{B} )}{B\,A^\frac{\beta-1}{2}
} \,I_{\beta-1}\Bigl(\frac{2\sqrt{A\,x}}{B}\Bigr) \\
& =& \left(\frac{x}{x^\prime \Theta}\right)^{\frac{\beta-1}{2}} \,\frac{
\exp(-\frac{x+x^\prime \Theta}{ B_0[1-\Theta]} ) }{B_0[1-\Theta]}
\,I_{\beta-1}\Bigl(\frac{2\sqrt{xx^\prime \Theta}}{B_0[1-\Theta]}\Bigr)
\,,  \label{Pfinal}
\end{eqnarray}
where $I_n(x)$ is the $n$th-order modified Bessel function of first kind~\cite{functionsBessel}.

In the limit $\gamma \Delta t>>1$, Eq.~(\ref{Pfinal}) tends to the
stationary PDF (\ref{limitP}), which is also obtained by performing the
integration $\int dx^{\prime }P\left( x,t|x^{\prime },t^{\prime }\right)
P_{s}\left( x^{\prime }\right) $. Still in the steady state, the two-time
joint PDF is
\begin{eqnarray}
&& P(x,t;x^{\prime },t^{\prime }) = P(x,t|x^{\prime },
t^{\prime})P_{s}(x^{\prime })   \\[3mm]
&=& \frac{ (x x^\prime)^\frac{\beta-1}{2} \exp( -\frac{x+x^\prime}
{B_0[1-\Theta]} ) } {\Gamma (\beta ) B_0^{\beta+1}[1-\Theta]
\Theta^\frac{\beta-1}{2} } I_{\beta-1}\left(\frac{2\sqrt{xx^\prime \Theta}}
{B_0[1-\Theta]}\right),  \label{joint2}
\end{eqnarray}
which in the long-term limit ($\Theta\to 0$), of course, becomes the
product of the stationary PDFs, $P_{s}( x) P_{s}( x^{\prime})$.

\subsection{Aggregation}
\label{aggregation}

Once known the propagator $P(x,t|x^{\prime},t^{\prime})$, given by Eq.~(\ref{Pfinal}),
and assuming stationarity and Markovianity, we can determine the $N$-time joint PDF
\begin{eqnarray}  \label{jointN}
&&P( x_1;\ldots;x_N) \equiv P( x_1,t_1;\ldots;x_N,t_N)   \\
&&= \prod\limits_{i=1}^{N-1}P\left(
x_{i+1},t_{i+1}|x_{i},t_{i}\right)\,P_{s}\left( x_1\right).
\end{eqnarray}
In Eq.~(\ref{jointN}), we can consider elements generated by Eq.~(\ref{sde})
that are equally spaced in time, such that $t_i=(i-1)\Delta t$, for $1\le
i\le N$. Then, we can evaluate the resulting stationary PDF of
$X=\sum\limits_{i=1}^{N} x_{i}$,
\begin{equation}
P_{N}\left( X\right) =\int \ldots \int \,dx_{1}\ldots dx_{N} \delta(X-\sum
x_{i})P\left( x_1;\ldots;x_{N} \right) \,.  \label{PN}
\end{equation}

For arbitrary $N\geq 2$, when $\Delta t>>1/\gamma $ ($\Theta \rightarrow 0$),
$P_{N}(X)$ must tend to the $N$-fold convolution of the Gamma
distribution (\ref{limitP}). Since it is closed under convolution, we have,
\begin{equation}
\lim_{\Delta t\rightarrow \infty }P_{N}\left( X\right) =\frac{X^{N\beta
-1}\exp (-\frac{X}{B_{0}})}{B_{0}^{N\beta }\Gamma (N\beta )}\,.  \label{PX_0}
\end{equation}

In the opposite limit $\Delta t<<1/\gamma $ ($\Theta \rightarrow 1$),
\begin{equation}
\lim_{\Delta t\rightarrow 0}P_{N}\left( X\right) =\frac{1}{N}P_{1}\bigl(
\frac{X}{N}\bigr)=\frac{X^{\beta -1}\exp (-\frac{X}{NB_{0}})}{
(NB_{0})^{\beta }\Gamma (\beta )}\,.  \label{PX_1}
\end{equation}

That is, in both limits, the sum of adjacent variables is Gamma distributed,
although with different values of the parameters. For intermediate degrees
of correlation, following the behavior of the Bessel factor for small and
large values (power-law and exponential, respectively)~\cite{bessel}, the
PDF of the sum is expected to remain close in shape to the Gamma
distribution, \textit{i.e.}, growing as a power-law at the origin and decaying
asymptotically with an exponential tail. The extreme cases suggest that as
the number $N$ of added variables accrues the power-law exponent increases
as well: the less, the more correlated the aggregated variables are.
Concomitantly, the exponential tail decays more slowly: the slower, the
larger the correlations.

\begin{figure}[t!]
\centering
\includegraphics*[bb=100 335 514 607, width=0.5\textwidth]{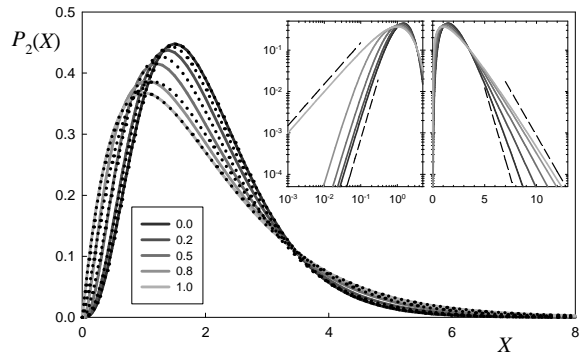}
\caption{PDF of the aggregation of two consecutive values, given by
Eq.~(\ref{P2}), for $(\beta,B_0)=(2,0.5)$ and different values of
$\Theta$ (hence $\Delta t$) indicated in the figure. Dotted lines
correspond to fittings with the Gamma distribution (for qualitative assessment), that exactly coincides
in the extreme cases $\Theta=0,\;1$. Insets show the PDFs in
logarithmic scales to better display the tails. Dashed lines are drawn for
comparison and correspond to $X^{\beta-1}$ and $X^{2\beta-1}$
in the left panel and to $\exp(-X/B_0)$ and $\exp(-X/(2B_0))$ in the
right-hand side one, which follow the asymptotic behaviors of the extreme
cases $\Theta=0$ and 1, respectively, while intermediate cases are
ruled by $\exp(-X/[(\sqrt{\Theta}+1)B_0])$. }
\label{fig:gammas}
\end{figure}

Let us analyze more carefully the particular case $N=2$. Probability distribution~(\ref{PN})
is explicitly,
\begin{equation}
P_{2}\left( X\right) =\frac{\sqrt{\pi }\;X^{\beta -1/2}\;\exp (-\frac{X}{
B_{0}[1-\Theta ]})I_{\beta -1/2}\Bigl(\frac{\sqrt{\Theta }X}{
B_{0}[1-\Theta ]}\Bigr)}{\Gamma (\beta )B_{0}^{\beta +1/2}[1-\Theta
]^{1/2}\,(2\sqrt{\Theta })^{\beta -1/2}}.  \label{P2}
\end{equation}
Taking into account the asymptotic behavior of the Bessel function, we can
make the following observations. For $\Theta \rightarrow 0$,
\textit{i.e.}, approaching independence, $P_{2}(X)\sim X^{\beta -1/2}\exp (-\frac{X}
{B_{0}[1-\Theta ]})X^{\beta -1/2}=X^{2\beta -1}\exp (-\frac{X}
{B_{0}[1-\Theta ]})$. For small enough $X$ the distribution goes to zero as
$P_{2}(X)\sim X^{2\beta -1}$ and exponentially decays with a characteristic
constant value equal to $B_{0}[1-\Theta ]$. This behavior changes when we
approach the full dependence case the functional form of which is
$P_{2}(X)\sim X^{\beta -1/2}\exp (-\frac{X}{B_{0}[1-\Theta ]})\exp
(\frac{\sqrt{\Theta }X}{B_{0}[1-\Theta ]})X^{-1/2}=X^{\beta -1}\exp
(-\frac{X}{B_{0}[1+\sqrt{\Theta }]})$. Therefore, the limit $\Theta \rightarrow 1$
yields for small $X$, $P_{2}(X)\sim X^{\beta -1}$ which despite being a
power-law has got a different exponent. The large $X$ decay keeps its
exponential form but with a different constant of decay equal to
$B_{0}[1+\sqrt{\Theta }]$. The behavior of Eq.~(\ref{P2}) between the two
limiting cases is exhibited in Fig. \ref{fig:gammas}.

It is noteworthy that for arbitrary $\Delta t$, distribution~(\ref{P2})
approaches the Gamma distribution (which is the exact distribution in the extreme cases)
at least for a few decades. On the one hand, the behavior near
the origin is the same as for $\Delta t\rightarrow \infty $. On the other hand,
an effective parameter for the exponential decay can be found to
adjust the tails since the true asymptotic behavior is attained after many
decades only.

\subsection{Correlations and moments}

The joint probability density given by Eq.~(\ref{joint2}) allows evaluation
of moments and two-time auto-correlation functions. To that end, it is
useful to calculate
\begin{equation*}
\tilde{C}_{nm}\equiv \langle x^{n}x^{\prime }{}^{m}\rangle \,=\,\int \int
dxdx^{\prime }P(x;x^{\prime })x^{n}x^{\prime }{}^{m}\,,
\end{equation*}
which is explicitly,
\begin{eqnarray} \nonumber
 &&\tilde{C}_{nm} =  \frac{\Gamma (\beta +m)\,\Gamma (\beta +n)B_{0}^{m+n}}
{\Gamma (\beta )^{2}}\times \\
& & [1-\Theta ]^{\beta +m+n}\,_{2}\mathcal{F}_{1}(\beta +m,\beta +n,\beta
,\Theta ).
\label{corr}
\end{eqnarray}

In particular, for any $\gamma\Delta t$, the (centered) linear
auto-correlation is (see also~\cite{yakovenko})
\begin{equation}  \label{c11}
C_{11}= \tilde{C}_{11}-\tilde{C}_{10} \tilde{C}_{01} = \beta\, \Theta\, B_0^2\,.
\end{equation}

From Eq.~(\ref{corr}), higher order correlations behave as
\begin{equation}
\tilde{C}_{nm} \simeq \frac{\Gamma( \beta+m )\Gamma(\beta+n) B_0^{m+n} }
{\Gamma(\beta)^{2}} \biggr(1-\frac{mn}{\beta} \Theta\biggr),
\end{equation}
for large $\gamma\Delta t$, at first order in $\Theta$. Meanwhile, for
small $\gamma\Delta t$, at first order in $1-\Theta$, one has
\begin{equation}
\tilde{C}_{nm} \simeq \frac{\Gamma( \beta+m+n ) B_0^{m+n} }{\Gamma(\beta)^{2}}
\biggr(1-\frac{mn(1-\Theta)}{\beta+m+n-1}\biggr).
\end{equation}
Hence, correlations of any order decay like a single exponential function,
with characteristic time $1/\gamma$.

However, a different behavior occurs if $\gamma$
fluctuates, within the framework that we will be considered in the following
section. Particularly, a power-law decay of the correlation function is
obtained if one assumes that the PDF of $\gamma$ decays exponentially~\cite{hugo-beck}.
In such a scenario it should be stressed that our stationary
distributions remain the same because they do not depend on parameter $\gamma $.

From Eq.~(\ref{corr}), the statistical raw moments can be directly obtained as
\begin{equation} \label{xn}
\langle x^{n}\rangle \,=\,\tilde{C}_{n0} \,=\,\frac{\Gamma(\beta +n)}
{\Gamma(\beta)}B_{0}^{n}.
\end{equation}

Hence the first centered moments are

\begin{eqnarray*}
\langle (x-\langle x\rangle)^2\rangle& =\beta B_0^2, \\[3mm]
\langle (x-\langle x\rangle)^3\rangle& =2\beta B_0^3.  \label{x23c}
\end{eqnarray*}

Concerning aggregation, for $N=2$, from Eq.~(\ref{P2}) and using the
properties in Ref.~\cite{functionsBessel}, statistical moments $\langle
X^{n}\rangle $ are given by
\begin{equation}
\left\langle X^{n}\right\rangle =\frac{\Gamma(2\beta +n) }{\Gamma (2\beta)}B_0^n [1-\Theta]^{\beta+n}\,_{2}\mathcal{F}_{1}(\beta +\frac{n}{2},\beta +\frac{1}{2}\left(
n+1\right) ,\beta +\frac{1}{2},\Theta).
\label{moments-sem-mix}
\end{equation}
The first raw moments read
\begin{eqnarray*}
\langle X\rangle&  =2\,\beta B_0, \\[3mm]
\langle X^2\rangle& =2\,\beta (2\beta+1+\Theta)B_0^2,   \\[3mm]
\langle X^3\rangle& =4\,\beta(\beta+1)(2\beta+1+3\Theta) B_0^3.  \label{X23}
\end{eqnarray*}
From where the first centered moments are
\begin{eqnarray}
\langle (X-\langle X\rangle)^2\rangle& =2\,\beta (1+\Theta)B_0^2, \\[3mm]
\langle (X-\langle X\rangle)^3\rangle& =4\,\beta (1+3\Theta)B_0^3.  \label{X23c}
\end{eqnarray}
Notice the increase in the moments with $\Theta$ due to the longer exponential tails.

For arbitrary $N$, in the limit of vanishing $\Theta$, the raw moments can be also obtained
from the moment generating function $M(z) = (1-B_{0}z)^{-N\beta}$, for any $N\ge 1$.

\section{Statistical mixing}

Let us now reckon that instead of constant, $B_{0}$ evolves stochastically,
independently of $x$, along the times eries of the primary process, for which
the time lag between successive points is $\Delta t$. In agreement with
previous observations~\cite{ca-vol}, $\eta =B_{0}^{-1}$ can be assumed Gamma
distributed with parameters $(\alpha ,1/\kappa _{0})$. Moreover, we
consider the update scale of $B_{0}$ (within which the parameter remains
basically constant) much larger than the timescale $1/\gamma $, so that the
average over different samples is obtained by performing the mixture,
\begin{equation}
P^{(M)}(\ldots )=\int d\eta P(\ldots |\eta )P(\eta ).  \label{mixing}
\end{equation}

Particularly, we can determine the mixed joint distribution
$P^{(M)}(x,t;x^{\prime },t^{\prime })$, which, after integration, reads
\begin{eqnarray}
&&P^{(M)}(x,t;x^{\prime },t^{\prime }) =\frac{\Gamma(\alpha +2\beta) }
{\Gamma(\alpha) \Gamma(\beta)^{2}} \frac{\kappa ^{\alpha }(xx^{\prime })^{\beta -1}
[1-\Theta]^{\beta}}{(\kappa +x+x^{\prime })^{\alpha +2\beta }}\times  \nonumber \\
&&{}_{2}\mathcal{F}_{1}\left( {\frac{\alpha }{2}+\beta ,\;\frac{1}{2}+
\frac{\alpha }{2}+\beta ,\;\beta ,}\frac{4xx^{\prime }\Theta}{(\kappa
+x+x^{\prime })^{2}}\right) \,,
\end{eqnarray}
where $\kappa =\kappa _{0}[1-\Theta]$. Subsequent integration over
$x^{\prime }$, allows us to obtain the global
distribution of $x$ (locally stationary) in the mixing process,
\begin{equation}
P^{(M)}(x)=\frac{\Gamma(\alpha +\beta) }{\Gamma(\alpha) \Gamma(\beta) }
\frac{\kappa _{0}^{\alpha } x^{\beta -1}}{(\kappa _{0}+x)^{\alpha +\beta }}
\,.  \label{Pmx}
\end{equation}

This PDF is a generalization of the Gamma distribution (recovered in the
limit $\alpha\to\infty$, while $\kappa_0/\alpha$ is kept finite), known as
$q$-Gamma. After suitable rescaling, it can also be cast into the form of an
$F $-distribution for which non-integer degrees of freedom are permitted.

It is worth noticing that, even in the limit $\Delta t>>1/\gamma $
(independence),
\begin{equation}
\lim_{\Delta t\rightarrow \infty }P^{(M)}(x;x^{\prime }) =
\frac{\Gamma(\alpha +2\beta) }{\Gamma( \alpha ) \Gamma ( \beta)^{2}} \frac{\kappa
_{0}^{\alpha }( x\,x^{\prime })^{\beta-1}} {(\kappa _{0}+x+x^{\prime
})^{\alpha +2\beta }}\,,  \label{Pmxx2}
\end{equation}
is different from the product $P^{(M)}(x)P^{(M)}(x^{\prime })$. Nonetheless, this does not
necessarily correspond to a case of correlated variables $x$ and $x^{\prime }$. Bearing in mind
the multi-variate Student-$t$ distribution~\cite{kotz}, we can understand our Eq.~(\ref{Pmxx2}) as a bi-variate
$F$-distribution of uncorrelated variables.

\subsection{Aggregation of mixed variables}

Performing the mixing operation given by Eq.~(\ref{mixing}) on Eq.~(\ref{P2}),
after integration, we obtain
\begin{eqnarray}
& P_{2}^{(M)}(X)=\frac{\Gamma (\alpha +2\beta )}{\Gamma (\alpha )\Gamma
(2\beta )}\frac{[1-\Theta ]^{\beta }\kappa ^{\alpha }X^{2\beta -1}}
{(\kappa +X)^{\alpha +2\beta }}\times   \nonumber \\
& \qquad {}_{2}\mathcal{F}_{1}\left( {\frac{\alpha }{2}+\beta ,\;\frac{1}{2}+
\frac{\alpha }{2}+\beta ,\frac{1}{2}+\beta ,}\frac{\Theta X^{2}}{(\kappa
+X)^{2}}\right) ,  \label{pmv}
\end{eqnarray}
which, in the independence limit $\Delta t>>1/\gamma $ (but still smaller
than any characteristic time of the mixing process), evolves into
\begin{equation}
\lim_{\Delta t\rightarrow \infty }P_{2}^{(M)}(X)=\frac{\Gamma (\alpha
+2\beta )}{\Gamma (\alpha )\Gamma (2\beta )}\frac{\kappa _{0}^{\alpha
}X^{2\beta -1}}{(\kappa _{0}+X)^{\alpha +2\beta }}\,.  \label{indep2}
\end{equation}
This result can be easily generalized to $N$ terms by iterated convolution
operations:
\begin{equation}
\lim_{\Delta t\rightarrow \infty }P_{N}^{(M)}(X)=\frac{\Gamma (\alpha
+N\beta )}{\Gamma (\alpha )\Gamma (N\beta )}\frac{\kappa _{0}^{\alpha
}X^{N\beta -1}}{(\kappa _{0}+X)^{\alpha +N\beta }}\,.
\label{indepN}
\end{equation}
On the other hand, in the opposite limit $\Delta t<<1/\gamma $
\begin{eqnarray}
\lim_{\Delta t\rightarrow 0}P_{N}^{(M)}(X) &=&\frac{1}{N}P^{(M)}
\bigl(\frac{X}{N}\bigr) \nonumber \\
&=&\frac{\Gamma (\alpha +\beta )}{N\,\Gamma (\alpha )\Gamma (\beta )}
\frac{\kappa _{0}^{\alpha }(\frac{X}{N})^{\beta -1}}{(\kappa _{0}+\frac{X}{N})^{\alpha +\beta }}\,.
\label{depN}
\end{eqnarray}
Notice that in both limiting cases a $q$-Gamma arises, although with
different exponents. This is a consequence of the mixing of the Gamma
distributions that rules the respective extreme cases.
For intermediate instances, the (effective) power-law exponent at the origin follows the same
scaling relation as the unmixed Gamma distribution. Meanwhile, the
exponent of the tail is insensitive to both $\Delta t$ and $N$,
conserving its value $\alpha +1$, which only depends on the degree of
inhomogeneities (given by $\alpha$), as it was empirically noticed in the
application given in Ref.~\cite{ca-vol}.

\begin{figure}[ht!]
\centering
\includegraphics*[bb=70 200 560 750, width=0.75\textwidth]{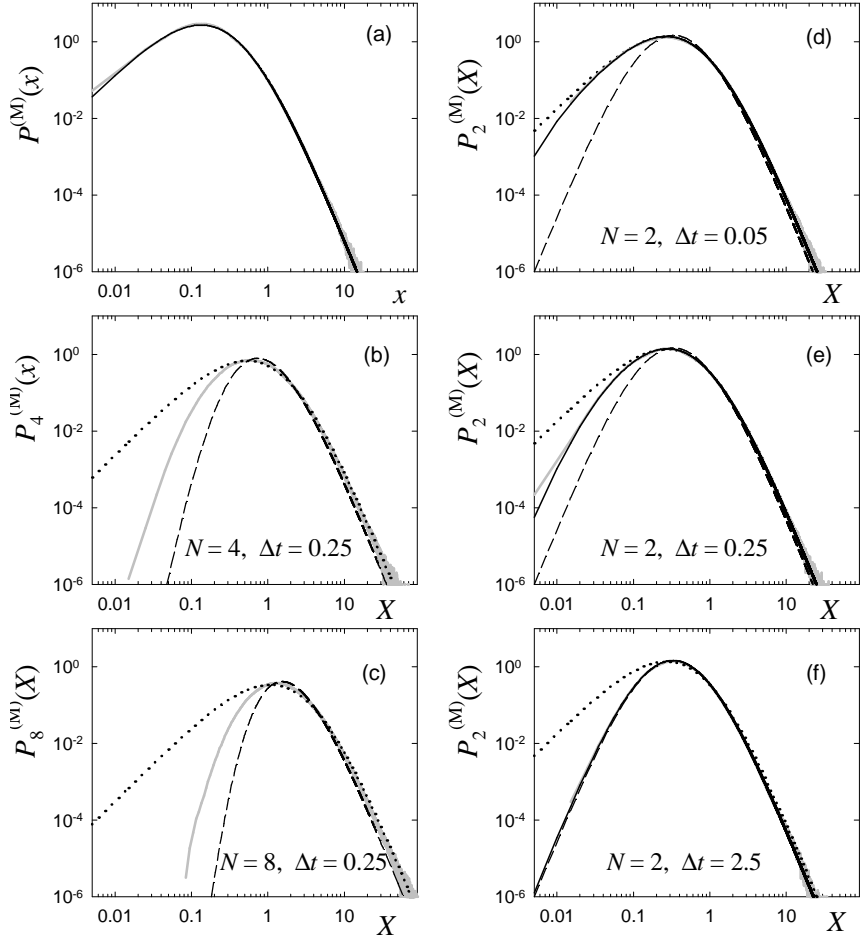}
\caption{PDFs of the mixing variable $x$ (a)
and agreggated variables $X$ for different values of $N$ and $\Delta t$ indicated
on the figure (b)-(f).
The (gray) lines join the points of the histograms built from numerical implementations of
Eq.~(\ref{sde}) with mixing.
Numerical integration of the stochastic differential equation was performed by means of an Euler algorithm,
with time step $10^{-3}$, and update of $1/B_{0}$, drawn from $\Gamma_{\alpha,1/\kappa_0}$,
was done at each $\delta t = 100 \, \gamma ^{-1}$.
Parameters values are $\gamma = 1$, $\alpha = 4$, $\beta = 3$, $\kappa _{0} = 1/3$.
In panel (a) the full line represents the  global PDF given by Eq.~(\ref{Pmx}).
Panels (b)-(f) exhibit the global PDF of the addition of $N=2$ (d-f), 4 (b) and 8 (c)
consecutive variables, at each $\Delta t$, for the same process.
For $N=2$, the full line represents Eq.~(\ref{pmv}, the dotted   line the
limit $\Delta t \rightarrow 0$ Eq.~(\ref{depN}) and the dashed  line the
independence limit given by Eq.~(\ref{indepN}).
All the plots are in  the same log-log scale for comparison.}
\label{fig:sim-pdf}
\end{figure}

These behaviors are illustrated in Fig.~\ref{fig:sim-pdf} for  numerical implementation of Eq.~(\ref{sde})
with mixing.
From panels (d) to (f) we verify that $P_{2}^{M}(X)$ departs from Eq.~(\ref{depN}) and approaches
Eq.~(\ref{indepN}), as $\Delta t$ swells and added variables become independent.
As aggregation proceeds, \textit{i.e.}, $N$ increases, the distribution shrinks below the maximum, departing
from the fully dependent case towards the independent one.
This is because  the exponent of the power law at the origin follows the behavior of the independent
case ruled by the $N\beta$ exponent (a reflection of the behavior at the origin of the unmixed case
as exemplified in Fig.~\ref{fig:gammas}), while as $X$ approaches the maximum, there is a crossover
towards the independence $\Delta t \to 0$ limit.
The figure also turns out evident that, contrarily,  the tail exponent does not depend on $N$
nor on $\Delta t$.  Its changeless value $\alpha+1$ indicates that the manifestation of inhomogeneities
stays invariant at the different accumulation scales.

Analogously to the Gamma approximation for the aggregation of
variables, discussed at the end of Section~\ref{aggregation}, the $q$-Gamma distribution
plays the same role after mixing.
As a consequence,  even when not the exact solution, the $q$-Gamma suits the PDF
generated by the mixed Feller diffusion.

\subsection{Correlations and moments after mixing}

As the term statistics of statistics suggests, the global statistical
properties correspond to the averaging of the locally stationary statistical
properties over the fluctuations. Therefore, central two-time correlation functions of $x$ are
\begin{eqnarray}
&&C_{nm}^{( M) } = \int \int dxdx^{\prime }d\eta P(x,x^{\prime })P(\eta)(x-\langle x\rangle)^{n}
(x^\prime-\langle x\rangle )^{m}\\
&&= \int d\eta P(\eta) C_{nm} .
\label{mixcnm}
\end{eqnarray}
Assuming that $P(\eta )$ follows a $\Gamma (\alpha ,1/\kappa _{0})$, from Eq.~(\ref{c11}),
we have
\begin{equation}
C_{11}^{\left( M\right) }
= \langle 1/\eta^2\rangle \beta \Theta =
\frac{ \beta \,\Theta\,\kappa _{0}^{2} }{(\alpha-1 )(\alpha-2) } ,
\end{equation}
which displays the same exponential decay as the unmixed process.

It is important to introduce two remarks. First, correlation functions and moments must
not be computed for time differences $\Delta t$ greater than the characteristic time
in which the fluctuating parameter can be considered constant.
Second, although we can feel enticed
to compute the overall (mixing)\ statistical properties by integrating the
variables using the $P^{( M) }( \ldots ) $ weights,
\textit{e.g.}, $C_{11}^{( M) }=   \langle x x^\prime\rangle^{( M)}
-\langle x\rangle^{( M) } \langle x^{\prime }\rangle^{( M) }$,
this is wrong, since one must keep in mind the non-stationary nature of
the stochastic process (due to the fluctuations in $\eta $).
Consequently, the statistical properties must be first computed locally and only
afterwards the parameter fluctuations taken into account.
Otherwise, spurious results may come forth such as for instance
centered correlations tending to a constant value different from zero in the long-time limit.
For instance, one has
\begin{equation}
\tilde{C}_{11}^{(M)} \equiv \langle x \, x^{\prime } \rangle ^{(M)} =
\frac{ \beta(\beta + \Theta)\kappa _{0}^{2}}{(\alpha-1)(\alpha -2) },
\label{covariance}
\end{equation}
which for $\Theta = 0$ is equal to
$\int \langle x \rangle \, \langle x^{\prime} \rangle P(\eta) \, d\eta$, as it should be according to the
discussion above.

In particular, moments after mixing must  be also computed by averaging over the statistics of
$\eta$, \textit{i.e.}, $\int (\ldots)P(\eta) d\eta $, the
expressions (\ref{xn})-(\ref{X23c}) obtained in the previous section for the
locally stationary process.

For $N=2$ aggregated variables, by integrating Eq.~(\ref{moments-sem-mix})
over $\eta$  with a $\Gamma(\alpha ,1/\kappa _{0})$ weight
(which leads to $\langle B_0^n\rangle =\kappa_0^n\Gamma(\alpha-n)/\Gamma(\alpha$) for $n<\alpha$),
one obtains
\begin{eqnarray*}
&&\langle X^{n}\rangle^{(M)} =
\frac{\Gamma( 2\beta +n ) \Gamma( \alpha -n) } {\Gamma( 2\beta ) \Gamma( \alpha )
}\kappa _{0}^{n}[1-\Theta]^{\beta +n}
\times  \\
&&\,_{2}\mathcal{F}_{1}(\beta +\frac{n}{2},\beta +\frac{1}{2} (n+1) ,\beta +
\frac{1}{2},\Theta).
\end{eqnarray*}
Notice that only moments with order $n<\alpha$ are defined.
The first raw moments (that can  also be obtained by
performing the mixing directly on Eqs.~(\ref{X23}))  are
\begin{eqnarray}
\langle X\rangle ^{(M)}& =\frac{2\,\beta \kappa _{0}}
{(\alpha -1)},  \\[3mm]
\langle X^{2}\rangle ^{(M)}& =\frac{2\,\beta (2\beta
+1+\Theta )\kappa _{0}^{2}}{(\alpha -1)(\alpha -2)},   \\[3mm]
\langle X^{3}\rangle ^{(M)}& =\frac{4\,\beta (\beta +1)
(2\beta +1+3\Theta )\kappa_{0}^{3}}{(\alpha -1)(\alpha -2)(\alpha -3)}. \label{X23mix}
\end{eqnarray}

Finally, for the centered moments, by integration of Eqs.~(\ref{X23c}) over
the fluctuations of $\eta$, one finds
\begin{eqnarray*}
\langle (X-\langle X\rangle)^2\rangle^{(M)}& =\frac{2\,\beta ( 1 +\Theta) \kappa _{0}^{2}}
{(\alpha -1)( \alpha -2) }, \\
& \\
\langle (X-\langle X\rangle)^3\rangle^{(M)}& =\frac{4\,\beta
(1 +3\Theta)\kappa _{0}^{3}  }{( \alpha -1) ( \alpha -2)( \alpha -3)}.
\end{eqnarray*}

Notice that, in the limit of $\alpha, \kappa_0 \to\infty$ with $\kappa_0/\alpha=B_0$ constant (that is
when the Gamma distribution of the fluctuating parameter tends to a Dirac $\delta$ centered at $1/B_0$), one
recovers the unmixed moments and correlations.

The case $\Theta\rightarrow 0$, \textit{i.e.}, the addition of independently
$F$-distributed variables ($N=2$) has been studied before from a pure
statistics perspective~\cite{Fvariates,Fvariates1,Fvariates2}. In these
cases the resulting distribution was approximated to the $F$-distribution by
imposing statistical moments matching. This procedure could be extended to
arbitrary $\Theta$ by means of the above expressions for the lowest order moments.

\section{Application to traded volume in financial markets}

Although the largest part of the work made on financial markets is devoted
to the (log-)price fluctuations and the volatility, it is recognized the
essential role of the traded volume for a trustworthy characterization of a
financial market global dynamics portrait~\cite{book}. As a matter of fact,
the price evolves in time when a certain quantity of equities is negotiated.

So far as we are aware, the first studies on high-frequency traded volume
were presented in~\cite{gopi-volume} wherein asymptotic power-law decay of
both the PDF and the auto-correlation have been held. Shortly after, another
study on the traded volume PDF was presented~\cite{obt}, but at that time,
the entire span of the traded volume values was taken into account and the
PDFs under analysis were very well adjusted by $q$-Gamma distributions. This
observation holds both for consolidated highly liquid stock
markets~\cite{smdq-vol,obt,volumes} (NYSE and NASDAQ) and for emerging ones like the
Brazilian \cite{ca-vol} and the Chinese~\cite{kertesz-zhou}. This fact
indicates universality of the functional form of the distribution function,
at least approximately, and therefore of the underlying dynamical mechanism
independently of the size of the market.

We shall now investigate the applicability of the dynamical scenario
presented in the preceding sections to model stock traded volumes. We
have analyzed two different paradigmatic examples: \emph{i)} the total volumes traded in
the emerging Brazilian stock market BOVESPA (a total of 9970 observations,
recorded at intervals of $30$ minutes, spanning the period from 3$^{rd}$
January 2005 to 13$^{th}$ September 2007) and \emph{ii)} the 1 minute records of
Pfizer (PFE) traded volume at New York Stock Exchange between the 1$^{st}$
July 2004 and 31$^{st}$ December 2004 in a total of 49585 registered values.
During the respective period each market can be considered in a regular state,
\textit{i.e.}, neither a crash nor other extreme behavior was earmarked.
Notwithstanding, future work should shed light on the traded volume dynamics and
its connection with the theory that predicts log-oscillatory behavior for the price~\cite{sornette,sornette1} in the advent of a crash.
It should be noted that crashes are empirically associated with herding phenomena
between agents and consequently with huge traded volumes~\cite{sornette2}.
Accordingly, the theory of log-oscillations must reflect the emergence of
a new dynamics enhancing large values of $\theta $ and vice-versa. Our assumption is supported by
prior empirical financial studies of daily time series which found an increase in trading volume over the previous
six months to a price plunge~\cite{ChenHongStein}.

\begin{figure}[b!]
\begin{center}
\includegraphics*[bb=150 240 500 660, width=0.5\textwidth]{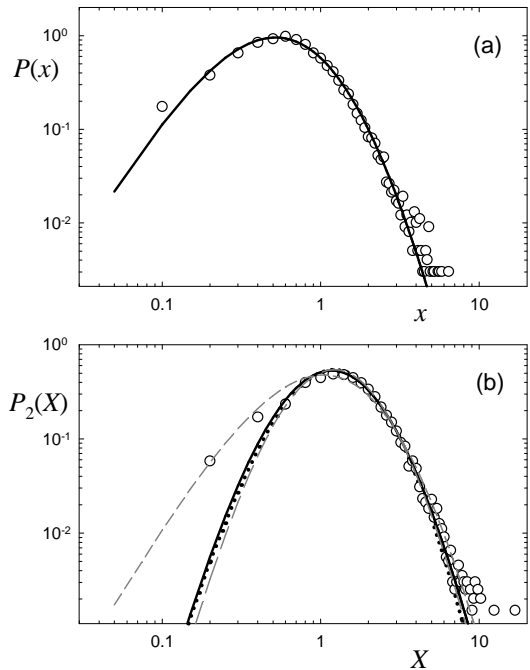}
\end{center}
\caption{(a) Empirical PDF of the BOVESPA $30$-minute
traded volume $x$ (circles)  and   best non-linear regression
result for Eq.~(\ref{Pmx}) (full line), yielding $(\alpha,\beta,\kappa_0)=(7.11,3.90,1.45)$.
(b) Empirical PDF of the BOVESPA $1$-hour traded volume, $X$,(cicles).
The full line corresponds to Eq.~(\ref{pmv}), $P_{2}^{(M)}(X=x+x^{\prime })$, with $(\alpha,
\beta,\kappa_0)$ as above and $\gamma =1.5$.
The dotted line represents a $q$-Gamma distribution with parameters
$(\alpha,\beta,\kappa_0) =(7.97,6.97,1.82)$ ($D_{KS}=0.014$)
obtained by a non-linear regression procedure.
For comparison, the (gray) short and long dashed lines correspond to the
limits of full dependence and independence, given by Eqs.~(\ref{depN})
and (\ref{indep2})), respectively.
}
\label{fig:bovespa}
\end{figure}

We start by determining the values of the set of parameters $\alpha$,
$\beta$, $\kappa _{0}$, by adjusting the traded volume empirical
distribution at the lowest time resolution of the data, that will be
considered the unit timescale, in each case.

For the Brazilian market, whose lowest scale is 30 min, the empirical
distribution of 30 min stock volumes (hence, $\Delta t=1$) is depicted
in Fig.~\ref{fig:bovespa}(a).
From the non-linear regression procedure (minimization of $\chi ^2$ error leading to the optimization of the parameters correlation matrix)we obtained
$(\alpha,\beta,\kappa_0) =(7.11 \pm 0.32, 3.90 \pm 0.04, 1.45 \pm 0.01)$ [the
respective Kolmogorov-Smirnoff distance, $D_{KS}$, to the empirical
probability distribution is equal to $0.016$, $\chi ^2 = 0.0380$, $R^2=0.994$]. Since the very small volume
regime ($x\lesssim 0.1$) is ruled by a different mechanism~\cite{ca-vol},
it was not considered in fitting procedures.

In order to model the empirical distribution of the 1-hour traded volume,
through Eq.~(\ref{pmv}), we used the values of the parameters resulting from
the numerical adjustment of the $30$-minute traded volume PDF, together with
the value of $\gamma$ obtained from the adjustment of the linear
auto-correlation function with an exponential decay. Although the
exponential does not describe well the correlations in long-term regimes, it
can be considered as a good approximation for the short timescales of
interest. From which we appraised $\gamma=1.5\pm 0.03$, since the
characteristic exponential time decay was $\tau=20$ min and we adopted $30$
min as the time unit.

Once obtained the values of parameters $(\alpha,\beta,\kappa_0,\gamma)$, we
compared the $1$-hour traded volume empirical PDF (symbols) with
Eq.~(\ref{pmv}) (full curve), as shown in Fig.~\ref{fig:bovespa}(b). We observe a fair
agreement between them ($D_{KS}=0.037$), specially recalling that very small volumes should not be
considered. However, notice that a simple
non-linear $q$-Gamma adjustment (dotted line) provides also a very good
description, with parameters $(\alpha,\beta,\kappa_0) =(7.97 \pm 0.35,6.97 \pm 0.06,1.82 \pm 0.01)$
[$D_{KS}=0.014$, $\chi ^2 = 0.0224$, $R^2=0.997$]
(dotted curve in Fig.~\ref{fig:bovespa}(b)). This illustrates once more how the $q$-Gamma model,
although approximate, appears to hold at different aggregation scales.
Furthermore, the fitting we present fails to reject the null hypothesis for the Pearson's statistical test with $P=0.05$.
Notice also that although correlations are not completely negligible at these timescales, they do not play an important
role in the resulting PDF, which is very close to the one that would be
obtained by assuming independence (gray long-dashed line).

It is remarkable that, volumes for a company in a developed market, recorded
at high frequency, as PFE, display the same qualitative features, despite correlations are
stronger at those (high frequency) time-scales (not shown, however, as illustration see Ref.~\cite{obt,volumes}).
In particular, from $q$-Gamma fitting, an almost constant value of  $\alpha$ is observed at
the different scales.
Moreover, PFE (in 2004), as well as the top 10 NASDAQ and NYSE stock volumes (in 2001) respectively
display $\alpha \simeq$ 4, 4 and 3, while $\alpha \simeq$ 7-8 for the Brazilian market, indicating
a lower degree of inhomogeneities in the former case. Then $\alpha$ constitutes an index
to detect and quantify the level of inhomogeneities of a market or period.

\section{Concluding remarks}

From the time-dependent PDF of the SDE describing mean-reverting square-root diffusion,
we derived the two-point (two-time) joint PDF in the steady state,
as well as the PDF of the addition of variables generated according to this
dynamical process. We further considered a
scenario in which the mean reverting term presents fluctuations that can be
introduced twofold: they  correspond to variations of the parameter
either over runs or within the same run in a time scale much larger than the
scale needed for the system to reach stationarity. Although our survey was
mainly inspired by previous empirical findings of inhomogeneities in the
traded volume flow corroborating a Gamma-Gamma phenomenological proposal, we
uphold that a similar approach might be applied on the study of systems
exhibiting inhomogeneous occurrence of Poisson events~\cite{emails} or simply on problems
for which the $q$-Gamma distribution has shown to be statistically relevant
like in granular media~\cite{kolb}. We have also discussed paradoxical results related to the non-commutativity
of averaging and mixing operations.

In both limits of full independence ($\Theta\rightarrow 0$) and full
dependence ($\Theta\rightarrow 1$), the PDF of the sum of consecutive variables
is a $q$-Gamma distribution. We  have shown that in intermediate situations, for arbitrary degree
of correlations, the upshot of aggregation  at different scales  is also well-described by that distribution,
although not the exact one. Moreover, while the increase at the origin is ruled by the independence behavior,
the tail is governed by the degree of inhomogeneities which manifest at any aggregation scale.
Then the $q$-Gamma form is approximately preserved.

We would like to stress that the model we have introduced explains why \textit{i)} even for $\Theta \neq 0$, the exponent $\beta$ of the associated PDF increases by increasing the number of added variables as well; \textit{ii)} the tail exponent, which may be considered
and indicator and quantifier of the presence of inhomogeneities, is preserved at different scales.

We have shown that all this scenario applies in the specific case of traded volume within periods of regular trading behavior (\textit{i.e.}, absence of bubbles/crashes) which represent the majority of the trading history. Nonetheless, for the fallout of stock market crashes, a further survey based on appropriate data that bridges the trading volume dynamics with the occurrence of extreme episodes (namely allowing for the theory of log-oscillations~\cite{sornette1}) is of manifest interest for an all-inclusive comprehension of financial markets dynamics. Especially, it would be interesting to understand the relation between the theory of log-oscillations and possible modifications in the description of the degree of inhomogeneities in a market (fluctuations in $\theta $) that are depicted by the parameter $\alpha $ during most of the trading records.

\ack

We acknowledge BOVESPA for furnishing the data. CA is grateful to
Brazilian agencies Faperj and CNPq for partial financial support and
SMDQ acknowledges funding by the European Union's Marie Curie Fellowship
programme.


\begin{thebibliography}{99}
\bibitem{bessel} The asymptotic behaviors of the modified Bessel function
are: $I_n(x) \to (x/2)^n/\Gamma(n+1)$, for $x<<\sqrt{n+1}$ and $I_n(x)\sim
e^x/\sqrt{2\pi x}$, for large $x>>n^2-1/4$.
\end{thebibliography}
\end{document}